\documentclass[fleqn,10pt]{SelfArx} 

\usepackage[english]{babel} 
\usepackage[super,compress]{cite}
\usepackage[justification=centerlast]{caption}[2020/09/12]

\definecolor{color1}{RGB}{0,0,90} 
\definecolor{color2}{RGB}{0,20,20} 

\usepackage{hyperref} 

\hypersetup{
	hidelinks,
	colorlinks,
	breaklinks=true,
	urlcolor=color2,
	citecolor=color1,
	linkcolor=color1,
	bookmarksopen=false,
	pdftitle={Title},
	pdfauthor={Author},
}

\JournalInfo{Condensed Matter -- Materials Science} 
\Archive{arXiv.org} 

\PaperTitle{Deterministic Covalent Organic Functionalization of Monolayer Graphene with 1,3-Dipolar Cycloaddition Via High Resolution Surface Engineering} 

\Authors{Luca Basta\textsuperscript{1}*, Federica Bianco\textsuperscript{1}, Aldo Moscardini\textsuperscript{1}, Filippo Fabbri\textsuperscript{1}, Luca Bellucci\textsuperscript{1}, Valentina Tozzini\textsuperscript{1}, Stefan Heun\textsuperscript{1}, Stefano Veronesi\textsuperscript{1}\ddag} 
\affiliation{\textsuperscript{1}\textit{NEST, Istituto Nanoscienze-CNR and Scuola Normale Superiore, Piazza S. Silvestro 12, 56127 Pisa, Italy}} 
\affiliation{*\textbf{Corresponding author}: E-mail: luca.basta1@sns.it} 
\affiliation{\ddag\textbf{Corresponding author}: Tel: +39 050 509882. E-mail: stefano.veronesi@nano.cnr.it} 

\Keywords{spatially-resolved organic functionalization --- monolayer graphene --- electron irradiation --- cycloaddition --- Raman spectroscopy --- AFM --- DFT} 
\newcommand{\keywordname}{Keywords} 

\Abstract{Spatially-resolved organic functionalization of monolayer graphene is successfully achieved by combining low-energy electron beam irradiation with 1,3-dipolar cycloaddition of azomethine ylide. Indeed, the modification of the graphene honeycomb lattice obtained via electron beam irradiation yields to a local increase of the graphene chemical reactivity. As a consequence, thanks to the high-spatially resolved generation of structural defects ($\sim$~100~nm), chemical reactivity patterning has been designed over the graphene surface in a well-controlled way. Atomic force microscopy and Raman spectroscopy allow to investigate the two-dimensional spatial distribution of the structural defects and the new features that arise from the 1,3-dipolar cycloaddition, confirming the spatial selectivity of the graphene functionalization achieved via defect engineering. The Raman signature of the functionalized graphene is investigated both experimentally and via ab initio molecular dynamics simulations, computing the power spectrum. Furthermore, the organic functionalization is shown to be reversible thanks to the desorption of the azomethine ylide induced by focused laser irradiation. The selective and reversible functionalization of high quality graphene using 1,3-dipolar cycloaddition is a significant step towards the controlled synthesis of graphene-based complex structures and devices at the nanoscale.}

\begin{document}

\maketitle 


\thispagestyle{empty} 


\section*{Introduction} 

The development of novel solid-state optoelectronic, sensing, and energy storage and conversion systems designed at the na\-no\-scale has enormously benefited from the recent advances in the science of nanomaterials. As an example, due to a higher active surface area, two-dimensional (2D) materials have shown promising results in terms of device integration density, compared to traditional 3D systems.\cite{Ferrari15,Bhimanapati15recent} In particular, since its discovery in 2004\cite{Novoselov04}, graphene has received extensive study as an ideal 2D candidate due to its outstanding mechanical\cite{Lee08,Koenig11}, optical\cite{Nair08}, and transport properties.\cite{Novoselov05,Berger06,Zhang05} Notably, graphene's excellent electrical mobility and conductivity,\cite{Du08,Bolotin08ultrahigh} as well as the low level of 1/f noise,\cite{Schedin07detection} makes it appealing for ultra-sensitive sensing applications.\cite{Yavari12graphene} Moreover, the 2D nature of graphene's structure allows to strongly influence its optoelectronical, catalytical, and gas storing characteristics by controlling the surrounding chemical environment. In fact, it has been shown that graphene's properties could be tailored by surface functionalization with suitable materials.\cite{Das08monitoring,Uddin14,Yu17,Ciammaruchi19,Yan21recent} In particular, the covalent functionalization of graphene via organic functional groups has been explored as a pivotal step towards the formation of graphene composites at the nanoscale.\cite{Bellucci20engineering} However, while graphene's high specific surface area of $2630$~$\mathrm{m^2/g}$ provides numerous possible binding sites,\cite{Ambrosi14} its chemical inertness makes it difficult to modify graphene's structure while preserving its exceptional properties.\cite{Mohan18}

A promising route in order to improve the reactivity of graphene is to introduce beneficial structural defects, i.e. engineered inhomogeneities and irregularities of the ideal structure.\cite{Jangizehi20defects} It is known that defects can profoundly influence the chemical properties of graphene,\cite{Liu15defects} as its mechanical, thermal, and electronic properties.\cite{Yang18structure} For example, defective graphene has shown increased chemical reactivity towards both diazonium salt-promoted radical addition reaction\cite{Ye17quantifying} and 1,3-dipolar cycloaddition reaction,\cite{Basta21covalent} which are commonly used approaches for the covalent organic functionalization of graphene-based materials.\cite{Criado15} While these approaches exploited already existing defects in graphene, like intrinsic boundaries and edges, a precise control on the spatial distribution of the defects would be more valuable in order to obtain a position-controlled surface functionalization. Nonetheless, defect engineering of graphene requires an accurate control of the amount of defects, and the minimization of contaminations from external sources remains critical.\cite{Zhou10making} Moreover, a precise control, with high lateral resolution, of the surface chemistry of graphene is fundamental for specific applications in band gap engineering, device fabrication, and sensors.\cite{Nourbakhsh10bandgap,Baraket12aminated,Yuan13graphene,Lee16defect}

Electron-beam irradiation (EBI), i.e. the exposure to a focused beam of energetic electrons, is a versatile method that satisfies the requirements for a controlled  introduction of structural defects in graphene. Indeed, depending on the electrons energy, a large variety of structural defects can be created, such as topological defects, vacancies, or sp\textsuperscript{3}-defects.\cite{Yang18structure} In order to ensure high chemical reactivity, dangling bonds or electron cloud deformations via bond rotation are necessary, and to fulfill this requirement, long exposure to low-energy electrons can be employed. Indeed, although the threshold beam energy for knockout in perfect graphene is 86 keV,\cite{Smith01electron} continuous irradiation with low-energy electrons results in the creation of point-like (single or double vacancies) and boundary-like defects.\cite{Malekpour16thermal,Lan14polymer,Sun15two} Consequently, by utilizing standard electron beam lithography systems, structural defects can be patterned across the graphene in a very flexible way. Recently, structural defects have been patterned by low-energy EBI in mechanically exfoliated monolayer graphene on silicon dioxide, with lateral resolution of few hundred nanometers.\cite{Basta21substrate}

Here, we present the selective covalent functionalization of defect engineered monolayer graphene with 1,3-dipolar cycloaddition (1,3-DC) of azomethine ylide. Atomic force microscopy (AFM) and Raman spectroscopy allow the analysis of the samples as-exfoliated, after the patterning, and after the functionalization. Raman spectroscopy maps show the appearance of the characteristic D peak, only in the patterned area, while AFM images confirm the spatial distribution of the pattern ($\sim$ 100 nm between the spots) designed via low-energy EBI. The 1,3-DC of azomethine ylide involves dipolarophile species (e.g. the localization of a C=C bond alkense of the graphene structure),\cite{Georgakilas02,Breugst20huisgen} which is favorable in presence of defects, hence introducing a selective control of the chemical reactivity of graphene. Indeed, the Raman analysis of the functionalized graphene flakes exhibits new features in the region 1050 - 1750 cm\textsuperscript{-1}, only in the patterned areas, whereas the unexposed areas still present the spectrum of pristine graphene, confirming the selectivity introduced via defects patterning. To deepen our understanding of the system, a model for functionalized graphene is built and ab-initio molecular dynamics, at density functional theory (DFT) level of theory, is exploited to evaluate the power spectrum (PS).\cite{thomas2013computing} Evaluating the PS for specific groups allows to identify the contribution of the functional groups of the azomethine ylide grafted on the graphene surface (methyl, carboxyl, and catechol groups) and of the modified vibrational modes of the graphene sheet. Furthermore, the functionalization is shown to be reversible under irradiation with a focused laser beam (100x objective, up to 1.6 mW). The desorption of the ylide is indicated by the recovery of the Raman spectrum towards the spectrum of non-functionalized patterned graphene, a result which opens the possibility for a controlled removing of the molecules and an even finer tailoring of the surface, toward a functional integrated circuit (IC) architecture. Indeed, laser ablation of azomethine ylide paves the way for a sequence of functionalization/ablation steps in order to build a functional super-array of molecules expressing different capabilities. This device processing mimics the IC design.

\section{Results and Discussion}

\addcontentsline{toc}{section}{Results and Discussion} 

\subsection{Patterned graphene flakes}
\addcontentsline{toc}{subsection}{Patterned graphene flakes} 

\begin{figure}[h]
	\centering
	\includegraphics[width=\linewidth]{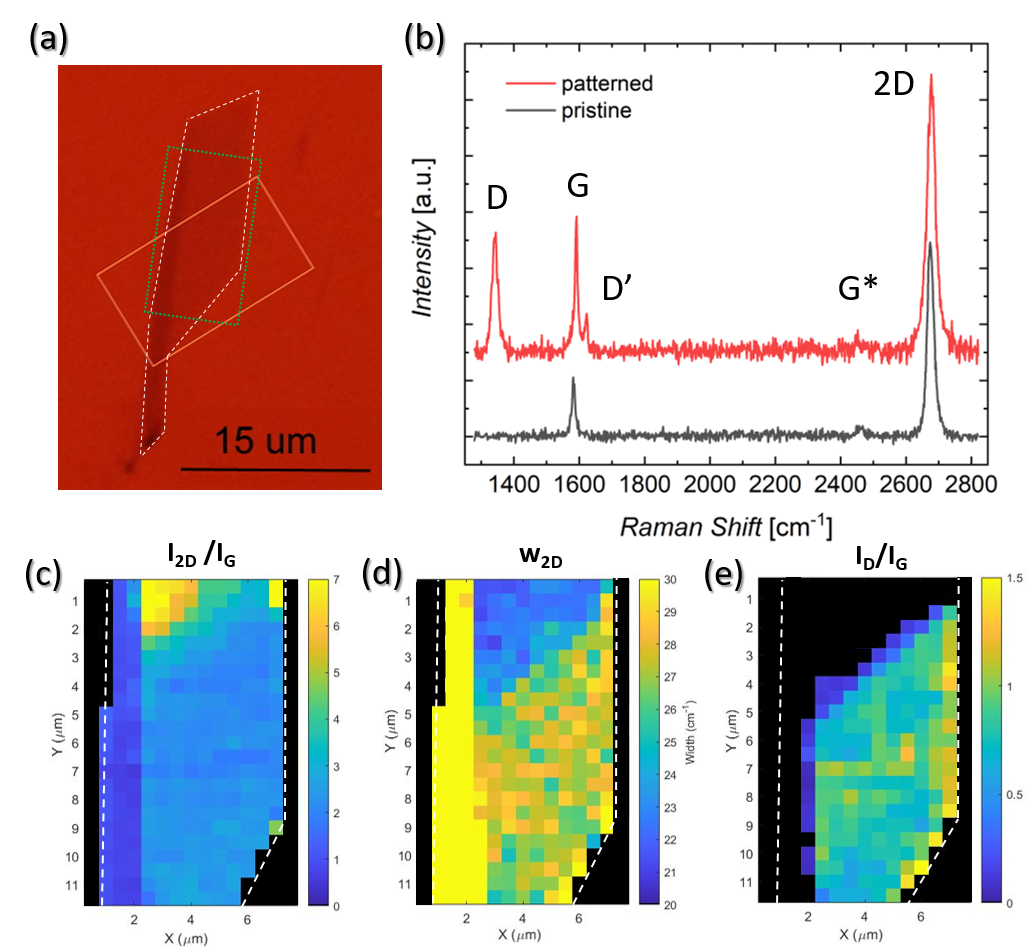}
	\small
	\caption{(a) Optical microscopy image of Flake 2. The white dashed line follows the edges of the flake, the orange rectangle indicates the patterned area (8 $\mu$m $\times$ 14 $\mu$m), and the green dotted rectangle is the area mapped by Raman spectroscopy. (b) Exemplary Raman spectra collected in the pristine (black) and patterned (red) areas of the flake (shifted in height), with the characteristic peaks labeled. Raman maps of (c) \textit{I}(2D)/\textit{I}(G) intensity ratio, (d) 2D peak width, and (e) \textit{I}(D)/\textit{I}(G) intensity ratio collected on the flake after exposure to EBI (in the area identified by the green dotted rectangle in panel (a)).}
	\label{fig:RamanMaps}
\end{figure}

Monolayer graphene flakes (labeled here as Flake 1, 2, and 3) are mechanically exfoliated on silicon dioxide substrates. Even if this was the first method used by Novoselov et al.\cite{Novoselov04} to produce graphene monolayers, mechanical exfoliation still plays a fundamental role in the rapid production of high quality single crystal graphene samples for research scale experiments.\cite{Huang20universal}
The pristine high quality monolayer graphene flakes are initially characterized by Raman spectroscopy and AFM. Then, a portion of each flake is exposed to EBI with electrons accelerated at 20~keV. The irradiated area is designed in order to expose half of the length  and the entire width of each flake (as shown in Figure~\ref{fig:RamanMaps}(a), where the exposed area on Flake 2 is 8~$\mu$m $\times$ 14~$\mu$m), with an electron beam step-size of 100 nm.

Raman spectroscopy is one of the most suitable techniques for the characterization of graphene-based materials, being able to provide both structural and electronic information, and allowing for fast and non-destructive measurements.\cite{Ferrari06,Ferrari13} Figure \ref{fig:RamanMaps}(b) shows a representative Raman spectrum collected on the unexposed area of the graphene monolayer (in black) along with a Raman spectrum recorded on the irradiated area (in red). As shown in the figure, the spectrum of the pristine graphene shows the characteristic G band, here centered at 1583 cm\textsuperscript{-1}, and 2D band, here centered at 2675 cm\textsuperscript{-1}. It is well-known that the G band comes from the in-plane bond-stretching motion of pairs of C atoms of the graphene ring, while the 2D peak is the second order of the D peak and originates from a double-phonon process where momentum conservation is satisfied. Since no defects are required for their activation, both G and 2D peaks are always present in the Raman spectra of graphene-based nanomaterials. The width (FWHM) of the 2D band is related to the number of layers of graphene sheets.\cite{Ferrari06} In particular, a single Lorentzian curve is the benchmark for the spectrum of single layer graphene.\cite{Malard09} Here, we can fit the 2D band with a single symmetrical Lorentzian curve, with a FWHM of 24.3 cm\textsuperscript{-1}. Besides the absence of the D peak, a further evidence for the high quality of the pristine monolayer graphene is the intensity ratio of the 2D and the G peaks, \textit{I}(2D)/\textit{I}(G) $\sim$ 3, for unexposed graphene. Finally, the weak G* band is visible, here centered at 2460~cm\textsuperscript{-1}, which arises from an intervalley process involving an in-plane transverse optical phonon and one longitudinal acoustic phonon (also called G + A\textsubscript{2U}).\cite{Krauss09}

The Raman spectrum of patterned graphene (the red line in Figure~\ref{fig:RamanMaps}(b)) exhibits additional bands, which are known to appear in the presence of structural defects. The D peak, here centered at 1342~cm\textsuperscript{-1}, and the D' peak, here centered at 1621~cm\textsuperscript{-1}, involve respectively intervalley and intravalley double resonance processes, and for low defect concentrations their intensities are proportional to the number of defects.\cite{Wu18} The intensity ratio of the D and D' peaks has been shown to indicate the nature of the defects in the graphene lattice. In particular, \textit{I}(D)/\textit{I}(D') reaches a maximum value of $\sim$~13 for sp$^3$ defects, decreases to $\sim$~7 for vacancy-like defects, and has a minimum of $\sim$~3.5 for boundary-like defects.\cite{Eckmann12} Here, we obtain \textit{I}(D)/\textit{I}(D')~$\sim$~4 for Flake 2 (\textit{I}(D)/\textit{I}(D')~$\sim$~5.5 for Flake 3), indicating a dominant presence of boundary-like defects. Acquiring Raman maps of the patterned flake, it is possible to verify the spatial distribution of the defects. In the patterned area, the intensity of the 2D band (Figure~\ref{fig:RamanMaps}(c)) decreases to an average value of \textit{I}(2D)/\textit{I}(G)~$\sim$~2, while the 2D band width (Figure~\ref{fig:RamanMaps}(d)) increases to an average value of FWHM $\sim$ 31~cm\textsuperscript{-1}. As defects are introduced in the graphene lattice, defect-dominated scattering processes become more likely, which results in a decrease of the intensity of the 2D peak and an increase in its width.\cite{Basko08theory,Eckmann12,Eckmann13raman} Mapping the intensity ratio of the D and G peaks (Figure~\ref{fig:RamanMaps}(e)) allows to confirm that no defects are introduced in the unexposed area of the flake. Moreover, from \textit{I}(D)/\textit{I}(G) $\sim$~0.9 it is possible to estimate the amount of disorder in the graphene sheet, both in case of boundary-like and point-like defects. For the former, we can estimate the nanocrystalline size \textit{L}\textsubscript{$\alpha$} using the Tuinstra-Koenig relation: \textit{I}(D)/\textit{I}(G) = \textit{C($\lambda$)}/\textit{L}\textsubscript{$\alpha$}, where \textit{C($\lambda$)} $\sim$~4.4~nm at $\lambda$ = 532~nm.\cite{Tuinstra70raman,Matthews99origin} Here, in case of patterned graphene, we obtain \textit{L}\textsubscript{$\alpha$} $\sim$~5~nm, confirming that we remain in the low-defect regime. In case of point-like defects in graphene, an useful measure for the amount of disorder is the distance between defects, \textit{L}\textsubscript{D}.\cite{Lucchese10quantifying} In case of low-disordered graphene, \textit{I}(D)/\textit{I}(G)$\propto$1/\textit{L}\textsubscript{D}\textsuperscript{2}. From the analysis presented in the work of Cancado et al.,\cite{Canccado11quantifying,Beams15raman} we can estimate {L}\textsubscript{D}~$\sim$~13~nm, which means a defects density {n}\textsubscript{D}~$\sim$~2$\cdot$10\textsuperscript{11}~defects/cm\textsuperscript{2}. Finally, the presence of defects influences also the G peak, which blue shifts to 1591~cm\textsuperscript{-1}, while its width remains almost constant (FWHM $\sim$ 12~cm\textsuperscript{-1}). Whereas the position of the G peak is particularly sensitive to doping, its width follows two competing mechanisms: due to doping it tends to decrease, while due to disorder it tends to increase. In the regime of a low concentration of defects (\textit{I}(D)/\textit{I}(G) $\leq$ 1) these two effects compensate each other.\cite{Tao13modification}

\begin{figure}[h]
	\centering
	\includegraphics[width=\linewidth]{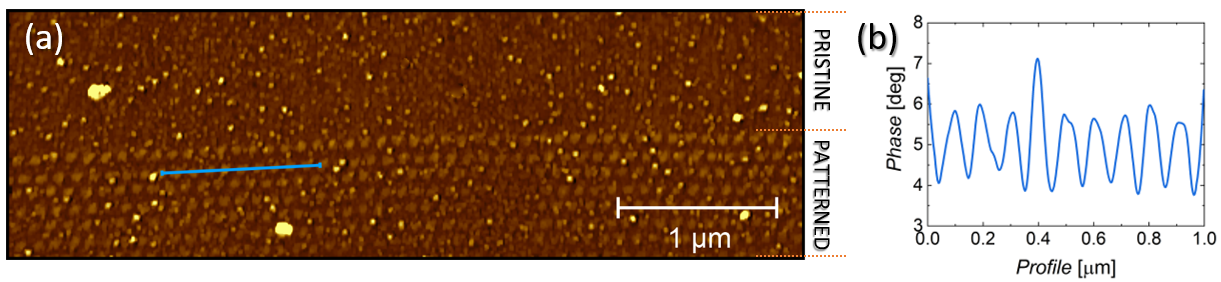}
	\small
	\caption{(a) AFM phase image of patterned graphene (Flake 1) showing the defect pattern after EBI exposure in the lower part of the image. (b) Profile taken along the blue line in panel (a), showing a step size of $\sim$ 100 nm.}
	\label{fig:AFMpattern}
\end{figure}

AFM is a powerful technique for the investigation of surfaces and interfaces at the micro- and nano-scale, allowing to reveal features with high spatial resolution (down to few nanometers). Here, the acquisition of AFM images of the patterned flake (shown in Figure~\ref{fig:AFMpattern}(a)) allows to clearly reveal the defect pattern. The defect spots are well-defined in the AFM phase channel, with a step size of $\sim$ 100 nm (see Figure~\ref{fig:AFMpattern}(b)), which exactly reproduces the input parameter of the EBI. Again, no defect pattern is visible in the unexposed area of the flake (see Figure~S1 of the Supporting Information for further AFM height and phase images).

\subsection{Functionalization of patterned graphene via 1,3-DC}
	
\addcontentsline{toc}{subsection}{Functionalization of patterned graphene via 1,3-DC} 
	
Patterned graphene is functionalized via a wet chemistry process, by adding N-methylglycine and 3,4-di\-hy\-drox\-y\-ben\-zal\-de\-hyde in 1-methyl-2-pyrrolidinone (NMP). The 1,3-DC of azomethine ylide occurs involving the localization of a C=C bond of the graphene lattice, as schematically represented in Figure~S2 of the Supporting Information. After the functionalization, the solvent is removed by several rinses (as described in detail in Section 4), and finally AFM and Raman spectroscopy are performed.

\begin{figure}[h]
	\centering
	\includegraphics[width=\linewidth]{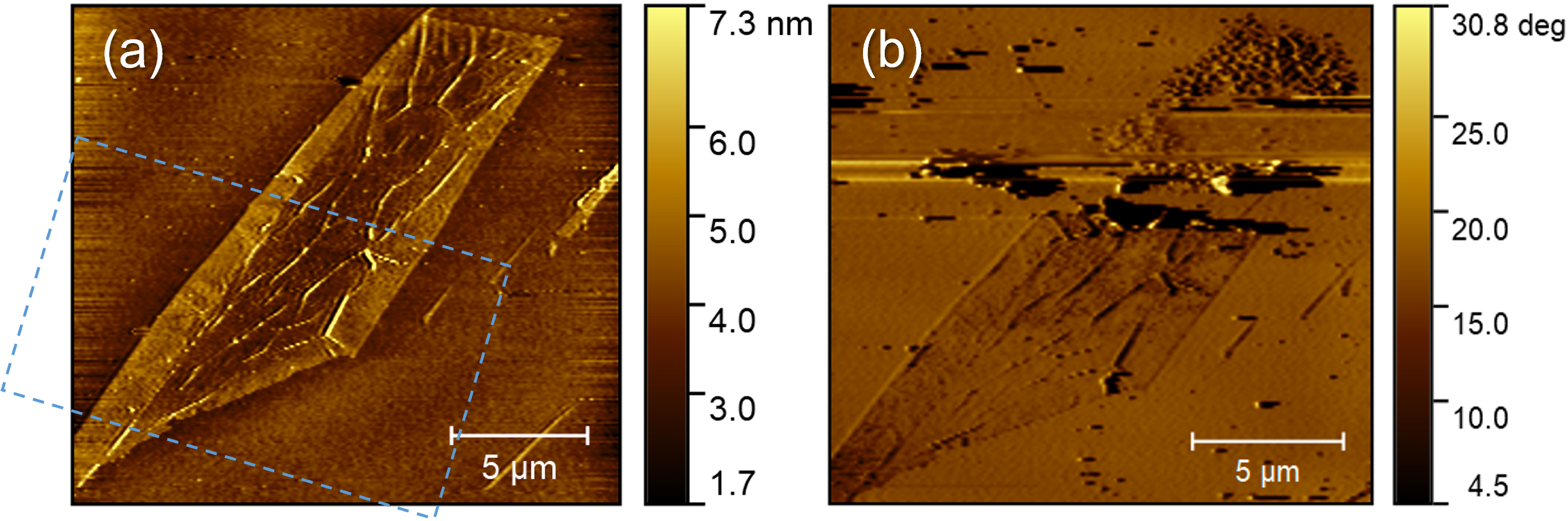}
	\small
	\caption{(a) AFM height image of graphene (Flake 2) after low-energy EBI (the blue rectangle shows the area which was exposed). (b) AFM phase image of graphene (Flake 2) after functionalization. The most part of the non-patterned zone of the flake detaches from the silicon dioxide substrate and folds, while the patterned zone remains adhered to the substrate.}
	\label{fig:AFMpostfunctionalization}
\end{figure}

Figure \ref{fig:AFMpostfunctionalization} shows AFM images of patterned graphene (Flake 2) before and after the functionalization (AFM images of Flake 3, before and after the functionalization, are shown in Figure~S3 of the Supporting Information). After the functionalization procedure, the part of the graphene flake that was exposed to EBI remains adhered to the silicon dioxide substrate, while most of the non-irradiated part of the flake detaches from the substrate and folds. NMP is an optimal organic solvent commonly used in 1,3-DC for its ability to favor the reaction by both dissolving the reagents and stabilizing the intermediates of reaction. On the other hand, it is among the best solvents for the dispersion of graphene, since its surface tension is very close to the ideal value of 40 mJ/m\textsuperscript{2}.\cite{Hernandez08} This means that NMP, while promoting the 1,3-DC of azomethine ylide on graphene, introduces an adhesion issue. Notably, patterned graphene exhibits an improved adhesion towards the silicon dioxide substrate (as can be seen in Figure~\ref{fig:AFMpostfunctionalization}(b)), overcoming the risk of dispersing the flake in the reaction solvent during the functionalizing procedure and allowing for further analysis and use of the graphene flakes after the functionalization in NMP. Notably, this result could help when fabricating devices based on graphene flakes functionalized with organic molecules. Indeed, designing defect patterns ad hoc would yield to both spatially resolved-fucntionalization and adhesion promotion of the functionalized graphene sheet.

\begin{figure}[h]
	\centering
	\includegraphics[width=\linewidth]{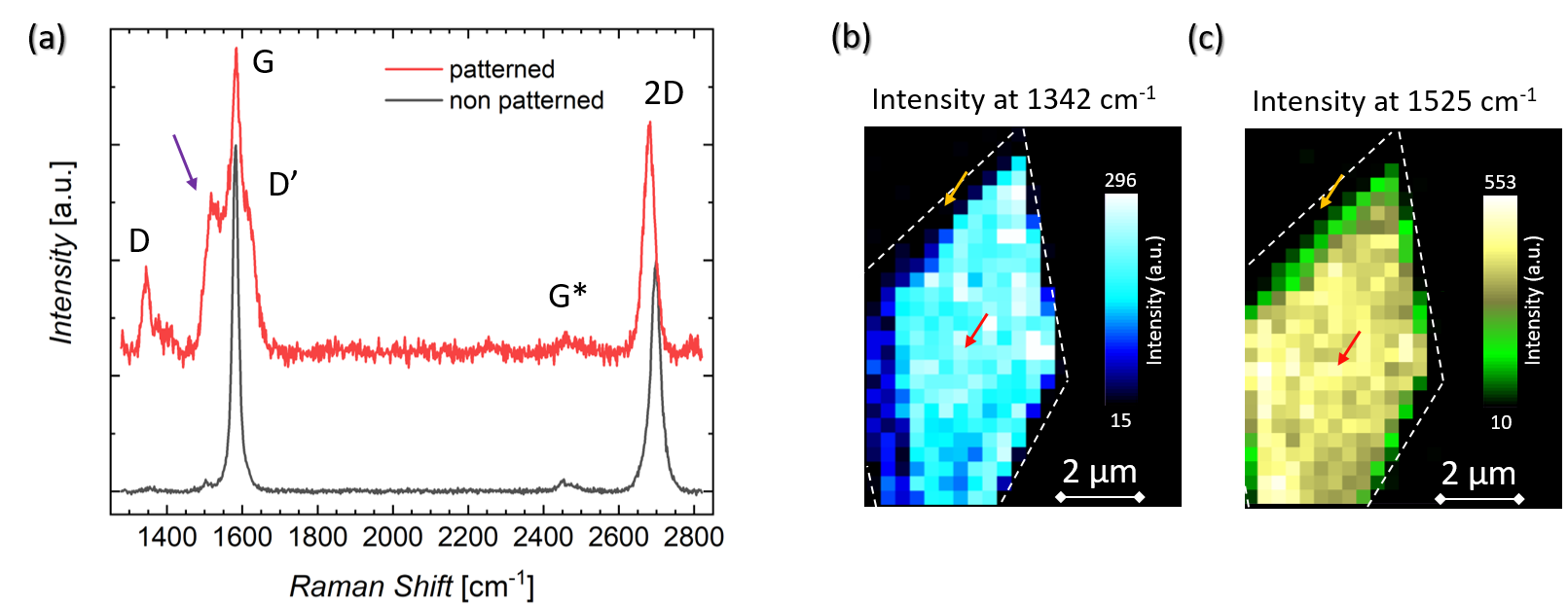}
	\small
	\caption{(a) Raman spectra of the functionalized Flake 2 collected in the non patterned (black line) and patterned (red line) areas of the flake (shifted in height). New Raman features are visible only in the spectrum of patterned functionalized graphene (the violet arrow indicates the most intense peak arising from the functionalization, at 1525 cm\textsuperscript{-1}). Raman maps of (b) D peak intensity and (c) intensity at 1525 cm\textsuperscript{-1} collected on Flake 2 after the functionalization procedure (the white dashed line follows the edges of the flake). The yellow and red arrows in panels (b-c) indicate respectively the positions where the spectra (shown in panel (a)) of non patterned and patterned graphene after functionalization were collected.}
	\label{fig:RamanMapsFunctionalized}
\end{figure}

The selectivity of the chemical functionalization is confirmed from the Raman maps collected on graphene after the functionalization procedure. Although most of the non-patterned area is folded, a narrow stripe of defects-free graphene is still present and its Raman spectrum can be investigated. The Raman spectra of functionalized graphene collected in the non-patterned area and in the patterned area of the flake show very different behavior (see Figure~\ref{fig:RamanMapsFunctionalized}(a)). The spectrum of functionalized patterned graphene after the functionalization procedure exhibits new features and modifications, while non-patterned graphene after the functionalization only presents a decrease in the intensity of the 2D peak in comparison to the non-patterned graphene spectrum acquired before the functionalization (\textit{I}(2D)/\textit{I}(G) passes from 2 to 0.65). In particular, there is a complete correspondence between the spatial distributions of the Raman intensity at 1342~cm\textsuperscript{-1} (shown in Figure~\ref{fig:RamanMapsFunctionalized}(b)), which corresponds to the characteristic defect activated peak (D peak) in graphene, and the Raman intensity at 1525~cm\textsuperscript{-1} (shown in Figure~\ref{fig:RamanMapsFunctionalized}(c)), which arises from the organic functionalization (as described in detail below). This correlation confirms the selectivity of the organic functionalization introduced by the spatially-resolved defect engineering of graphene via EBI.

\begin{figure}[h!]
	\centering
	\includegraphics[width=0.9\linewidth]{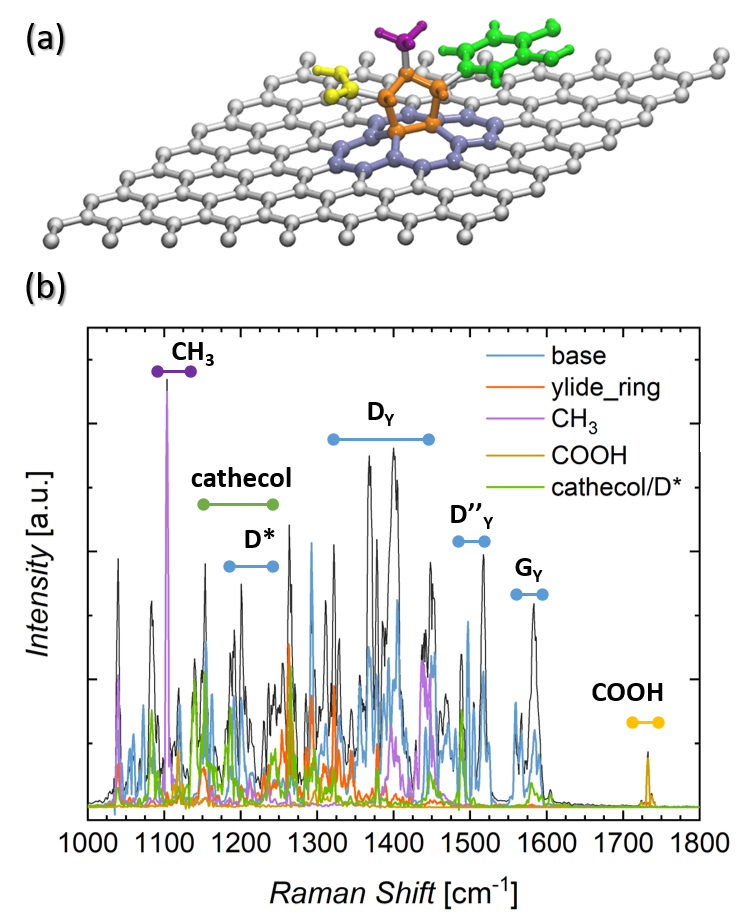}
	\caption{(a) Model of the azomethine ylide attached to graphene. The functional groups of interest are highlighted: carboxyl group = yellow, pyrroline ring = orange, methyl group = violet, cathecol group = green. The carbon atoms of graphene at the base of the molecule are highlighted in blue. (b) Power spectrum (PS) of the velocity autocorrelation function (black). Projections on the PS for the functional groups of interest are highlighted in the same colors of panel (a). The regions that correspond to the peaks detected in the Raman spectra are indicated by colored lines and labeled. It is worth to recall that the intensities of the peaks in the simulated PS do not directly correspond to the ones from the experimental Raman spectra, where additional selection rules are involved.}
	\label{fig:SimSpectrum}
\end{figure}

In order to deepen our understanding of the functionalized graphene after 1,3-DC of azomethine ylide and assign the new Raman bands that arise, a model for functionalized graphene is developed (see Figure~\ref{fig:SimSpectrum}(a)). The system is subjected to 38 ps of ab-initio  molecular dynamics (MD) simulation at 300~K. During the molecular dynamics simulation, the lattice does not undergo a large conformational rearrangement. The vibrational density of states is obtained as the Fourier transform of the velocity autocorrelation function\cite{thomas2013computing}. The power spectrum (PS) is then decomposed and analyzed by computing the power spectra of the autocorrelation function of appropriate groups of atomic coordinates. The functional groups of interest are highlighted in different colors in Figure~\ref{fig:SimSpectrum}(a)), and their corresponding projections are shown in Figure~\ref{fig:SimSpectrum}(b).

\begin{figure}[h]
	\centering
	\includegraphics[width=0.9\linewidth]{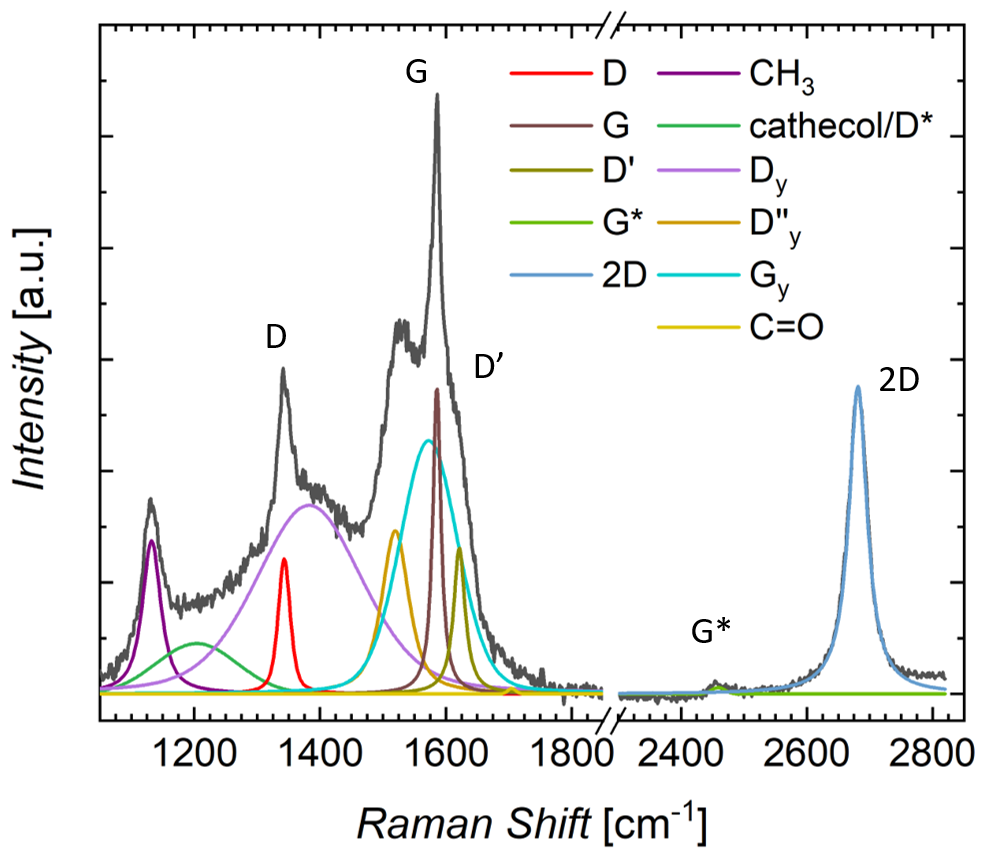}
	\small
	\caption{Raman spectrum of functionalized patterned graphene (Flake 3) collected at 120 $\mu$W laser power (the parts of the spectrum before and after the break were collected in different measurements, shown in Figure~S5 of the Supporting Information). A fit to the data is also shown, and the main peaks of graphene are labeled.}
	\label{fig:Ramanpostfunctionalization}
\end{figure}

Figure \ref{fig:Ramanpostfunctionalization} presents in detail the experimental Raman spectrum of functionalized patterned graphene collected on Flake 3 and its fit (the spectrum and its fit for Flake 2 is shown in Figs. S4 and S6 of the Supporting Information). Looking at the 2D peak, here centered at 2682 cm\textsuperscript{-1}, the intensity ratio \textit{I}(2D)/\textit{I}(G) decreases from 2 to a value $\sim$ 1.5, and the 2D width increases to a FWHM $\sim$ 37 cm\textsuperscript{-1}. Also the intensity ratio between the D peak, here centered at 1343 cm\textsuperscript{-1}, and the G peak, here centered at 1586~cm\textsuperscript{-1}, decreases from an initial value of \textit{I}(D)/\textit{I}(G) $\sim$ 1, for patterned graphene, to a value of \textit{I}(D)/\textit{I}(G) $\sim$ 0.3, for patterned functionalized graphene. The decrease of \textit{I}(D)/\textit{I}(G) can be explained considering that the molecules of azomethine ylide are grafting onto graphene's most favorable bonding sites in the patterned area, which are close to the defects, possibly leading to a local structural relaxation and a decrease in the Raman intensity of the defects themselves, as already seen in previous works.\cite{Basta21covalent} A further sign of the functionalization is the rise of new bands in the Raman shift region 1050 - 1750 cm\textsuperscript{-1}, which can be assigned with the aid of the computed PS (see Figure~\ref{fig:SimSpectrum}(b)).

The characteristic vibrational peak from C=O is centered at 1730~cm\textsuperscript{-1} (as shown in the PS) and would reasonably arise from the carboxyl group (COOH) of the azomethine ylide is visible. This corresponds to the peak centered at 1705~cm\textsuperscript{-1} in the experimental Raman spectrum, as also seen in literature.\cite{Basta21covalent,Meade10}
Another functional group of the azomethine ylide is the methyl group (CH\textsubscript{3}), which possesses characteristic vibrational modes around 1110~cm\textsuperscript{-1} (as shown in the PS). This could correspond to the new band revealed in the Raman spectrum of functionalized graphene, centered at 1130~cm\textsuperscript{-1}, which is, therefore, labeled CH\textsubscript{3}.

The cathecol group of the azomethine ylide, being a more complex part of the molecule, possesses a broader set of vibrational stretching modes, which fall around 1150 - 1250~cm\textsuperscript{-1} (as shown in the Figure~\ref{fig:SimSpectrum}(b)). Moreover, in the same region also contributes the D* band, usually found around 1180 - 1200~cm\textsuperscript{-1}, which can be related to disordered graphitic lattices provided by sp\textsuperscript{2}-sp\textsuperscript{3} bonds at the edges of networks.\cite{Sadezky05} These bands correspond to the wider band in the Raman spectrum that is centered at 1204~cm\textsuperscript{-1}, and is labeled as catechol/D*. Finally, three sets of vibrational modes can be identified in the region between the D and the G bands. From the projections on the PS of the carbon atoms of graphene at the base of the azomethine ylide, it can be inferred that these bands originate from vibrational normal modes of the modified graphene lattice. In particular, these bands are very close in frequency to the well-known D, D", and G bands of graphene. The D" band, usually seen in the range 1500 - 1550~cm\textsuperscript{-1}, is thought to be related to either the amorphous phase (increasing with the decrease of crystallinity)\cite{Vollebregt12} or to interstitial defects associated with the functionalization with small molecules.\cite{Sadezky05,Goodman13,Claramunt15} In the case of functionalized graphene, the presence of a molecule of azomethine ylide grafted on the graphene sheet breaks its homogeneity and slightly modifies the symmetry and the frequency of the vibrational normal modes of the graphene lattice. For example, computing the normal modes of the functionalized graphene, in the region around the G and D peaks, vibrational symmetries very similar to the ones of the canonical G and D peaks appear (see Figure~S7 of the Supporting Information). Hence, the bands in the Raman spectrum centered at 1383~cm\textsuperscript{-1}, at 1520~cm\textsuperscript{-1}, and at 1574~cm\textsuperscript{-1} are labeled here as D\textsubscript{Y}, D"\textsubscript{Y}, and G\textsubscript{Y} respectively, since they originate from modified graphene lattice vibrations due to the grafting of azomethine ylide.

In comparison with the results shown in a previous work (see ref. \citen{Basta21covalent}), thanks to the high quality of the graphene substrate, here it is possible to identify the Raman peaks arising from the azomethine ylide in the region 1000 - 1300 cm\textsuperscript{-1}, which is a region where several bands related to structural defects appear, especially in highly defected graphene samples. On the other hand, the second order bands of the stretching vibrational bands of the ylide (which should arise in the region 1900 - 2400 cm\textsuperscript{-1}) here are not detectable, possibly due to a lower degree of functionalization (because of a lower presence of defects in the graphene substrate).

\begin{figure}[h]
	\centering
	\includegraphics[width=0.9\linewidth]{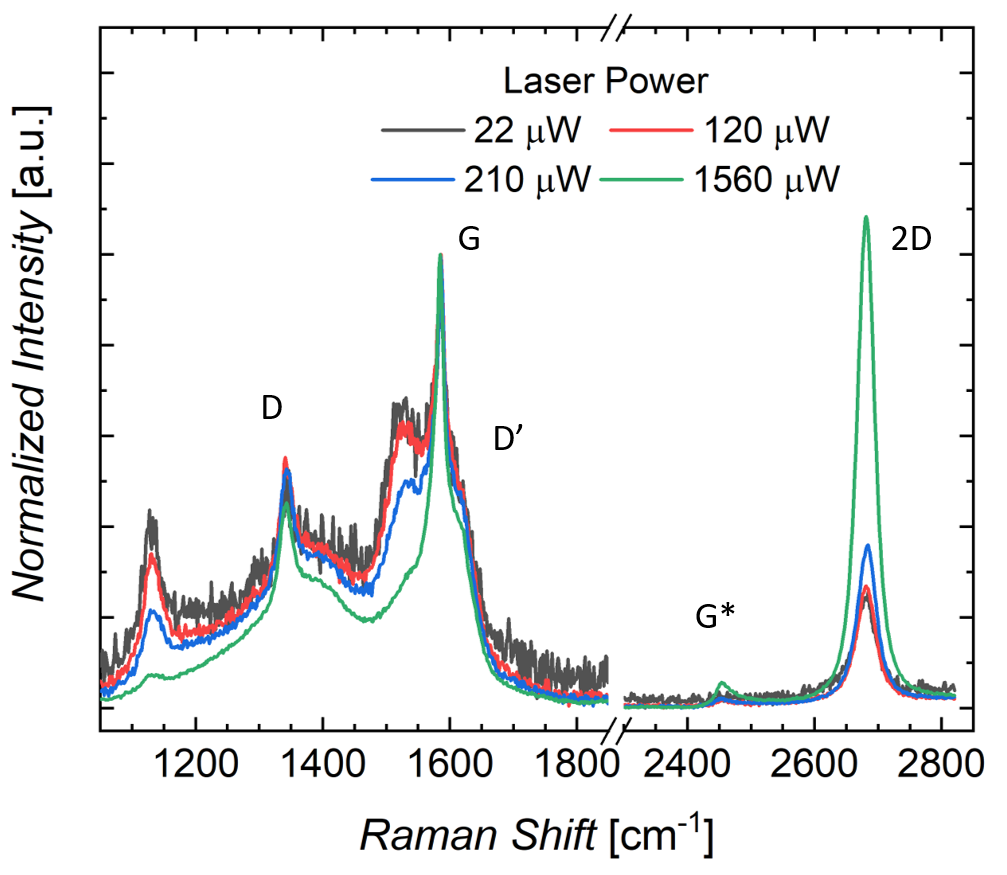}
	\small
	\caption{Raman spectra (normalized to the intensity of the G peak) of functionalized graphene (Flake 3) collected at increasing laser power (the parts of the spectra before and after the break were collected in different measurements, shown in Figure~S8 of the Supporting Information).}
	\label{fig:Ramandesorption}
\end{figure}
	
\subsection{Reversibility of the functionalization}
	
\addcontentsline{toc}{subsection}{Reversibility of the functionalization} 
	
Finally, the evolution of the surface functionalization for increasing laser power irradiation is investigated. In order to explore it, subsequential Raman spectra are collected from the same position in the functionalized patterned area of the graphene flake (Flake 3). The power of the excitation laser can be varied with different internal optical filters allowing measurements at various incident powers, which were accurately measured with a power meter. The actual incident laser powers used in this work are 22 $\mu$W, 120 $\mu$W, 210 $\mu$W, and 1560 $\mu$W. As shown in Figure~\ref{fig:Ramandesorption}, for increasing laser power irradiation, the intensity of the Raman modes previously assigned to the functionalization of the graphene lattice with azomethine ylide gradually decreases. In order to quantify the modifications in the Raman spectra, fits with the same bands that were identified from the previous analysis (see Figure~\ref{fig:Ramanpostfunctionalization} for reference) were performed. For each Raman spectrum, it is useful to normalize all peak intensities to the intensity of the G peak (see Table \ref{tab:Ramanratio}). Looking at the bands that originate from the functional groups of the azomethine ylide, we notice that \textit{I}(CH\textsubscript{3})/\textit{I}(G) passes from 0.83 (for the spectrum collected at 22 $\mu$W) to 0.07 (for the spectrum collected at 1560 $\mu$W). Similarly, \textit{I}(catechol/D*)/\textit{I}(G) passes from 0.18 to 0.10. Also the modified graphene vibrational bands gradually decrease in intensity when increasing the incident power. In fact, \textit{I}(D\textsubscript{Y})/\textit{I}(G) passes from 0.83, for the spectrum collected at 22 $\mu$W, to 0.39 for the spectrum collected at 1560 $\mu$W. Likewise, \textit{I}(D"\textsubscript{Y})/\textit{I}(G) decreases from 0.64 to 0.05, and \textit{I}(G\textsubscript{Y})/\textit{I}(G) decreases from 1.28 to 0.42. Finally, \textit{I}(2D)/\textit{I}(G) notably increases from 0.88 to 1.54, while its FWHM decreases from 40~cm\textsuperscript{-1} to 30~cm\textsuperscript{-1}. The partial recovering towards the spectrum of non-functionalized patterned graphene suggests the desorption of the azomethine ylide from the surface. Often, surface reactions promoted by laser irradiation are thermally activated processes, but in this case the substrate heating is estimated to be lower than a few degrees, due to the very low laser power used in the Raman measurements. Therefore, other photoinduced chemical processes are involved in the desorption of the azomethine ylide. Possibly, the energy of the incident photons (2.3 eV) would activate a resonant vibrational excitation\cite{Chuang85photodesorption} that can overcome the two C-C bonds of the molecule with the graphene surface (the C-C bond energy is estimated to be $\sim$ 1 eV).\cite{Denis10} Consequently, a photoinduced dissociation process could activate the retro-cycloaddition (as seen in case of functionalized fullerenes),\cite{Delgado11mass} resulting in the desorption of the azomethine ylide. The complete understanding of the desorption process still requires further investigation, however, our results undoubtedly indicate the reversibility of the functionalization. Indeed, this is a valuable result with a view to designing a more complex surface functionalization or simply to recovering a clean graphene sheet.

\begin{table*}[h!]
	\small
	\caption{Intensity of the Raman peaks arising from functionalized graphene normalized to the G peak intensity, for spectra collected at different incident laser power.}
	\label{tab:Ramanratio}
	\begin{tabular*}{\textwidth}{@{\extracolsep{\fill}}cccccccc}
		\hline
		laser power ($\mu$W) & \textit{I}(CH\textsubscript{3})/\textit{I}(G) & \textit{I}(cathecol/D*)/\textit{I}(G) & \textit{I}(D)/\textit{I}(G) & \textit{I}(D\textsubscript{Y})/\textit{I}(G) & \textit{I}(D"\textsubscript{Y})/\textit{I}(G) & \textit{I}(G\textsubscript{Y})/\textit{I}(G) & \textit{I}(2D)/\textit{I}(G) \\
		\hline
		22 & 0.83 & 0.18 & 0.37 & 0.83 & 0.64 & 1.28 & 0.88 \\
		120 & 0.50 & 0.17 & 0.44 & 0.62 & 0.54 & 0.83 & 0.92 \\
		210 & 0.28 & 0.15 & 0.44 & 0.56 & 0.30 & 0.77 & 0.98 \\
		1560 & 0.07 & 0.10 & 0.30 & 0.39 & 0.05 & 0.42 & 1.54 \\
		\hline
	\end{tabular*}
\end{table*}
	
\section{Conclusions}
	
\addcontentsline{toc}{section}{Conclusions} 
	
In this work we have presented a technique that successfully allowed a spatially-resolved functionalization of monolayer graphene by combining low-energy EBI with 1,3-dipolar cycloaddition of azomethine ylide. Low-energy EBI is shown to be a versatile method for defect engineering graphene, allowing for the introduction of mild modifications of the graphene lattice. Due to the defects pattern, which can be designed with the high spatial resolution of the EBI (few nm), it was possible to selectively enhance both the chemical activity of the graphene sheet towards the organic functionalization and the adhesion of the graphene monolayer to the silicon dioxide substrate. AFM and Raman spectroscopy analysis demonstrated a homogeneous averaged distribution of the defects in the patterned area after low-energy EBI. After the functionalization procedure, only the patterned area of graphene exhibited new Raman features, which were assigned with the aid of DFT simulations of the vibrational power spectrum of functionalized graphene. These new features originated from the presence of the azomethine ylide, both from the functional groups of the molecule itself and from the modifications that the molecule induced in the vibrational normal modes of the graphene sheet. Finally, the organic functionalization was shown to be stable but reversible through laser irradiation.

Whereas cycloaddition reactions were already used in previous works for the covalent functionalization of graphene in the liquid phase, the achievement of a selective organic functionalization of a higher quality graphene system like exfoliated monolayer flakes opens the route for a wider range of applications. In fact, by exploring the controlled use of defect engineering, the precise building of specific nanostructures is possible, such as electrochemical devices for sensing or gas storage. Moreover, for example, specifically designed surface patterns will support the assembly of graphene multilayer systems, where the molecules that act as spacers or linkers can be precisely positioned over the tailored graphene surface. These results also validate the interest in further exploring the nature of the defects, towards the perspective of designing printed 2D integrated circuits, in which the reversibility allows for the recovery of the initial system or a further tailoring of the surface.


\section{Methods}

\addcontentsline{toc}{section}{Methods} 

\subsection{Chemicals}
1-methyl-2-pyr\-rol\-id\-i\-none (ReagentPlus, 99\%), dichloromethane (puriss, $\geq$99.9\%) ethanol (puriss, $\geq$96\%), N-methylglycine (98\%), and 3,4-dihydroxybenzaldehyde (97\%) were purchased from Sigma-Aldrich/Merck.

\subsection{Graphene flake exfoliation}
Graphene flakes were micromechanically exfoliated from highly oriented pyrolytic graphite on boron-doped Si substrate with 300-nm-thick thermally grown SiO\textsubscript{2}. Metallic markers were lithographed by electron-beam for defining flake positions. Before the exfoliation process, the substrate surface was cleaned by e-beam resist residue removal solution and by oxygen plasma at 100 W for 5 minutes.

\subsection{Graphene flake patterning}
The defective areas were created by irradiating --- in a single step --- the graphene flake with electrons accelerated to 20 keV. The electron-beam was scanned with a step-size of 100~nm, with a current of about 0.15 nA. The dose was 40 mC/cm\textsuperscript{2}, resulting in a dwell-time of 30 ms.

\subsection{1,3-DC of azomethine ylide on graphene flake}
To perform the organic functionalization, 3,4-\-di\-hy\-drox\-y\-ben\-zal\-de\-hyde (40 mg, 0.29 mmol) and an excess of N-methylglycine (40 mg, 0.45 mmol) were added in 20~mL of NMP, in a Schlenk-type glass flask. The chip with the graphene flakes was submersed in the reaction mixture, placed on a custom made Teflon support, which allowed for magnetic stirring underneath. The reaction environment was kept at 150 °C for 16 h, under magnetic stirring. In order to limit secondary reactions from the oxidation of the solvent at high temperature, an inert atmosphere (N\textsubscript{2}) was kept during the functionalization reaction. The chip was then thoroughly washed several times with clean NMP, ethanol, dichloromethane, and finally air dried.

\subsection{Characterization Techniques}
Raman spectroscopy was carried out with a Renishaw InVia system, equipped with a confocal microscope, a 532 nm excitation laser and a 1800 line/mm grating (spectral resolution 2 cm\textsuperscript{-1}). Single spectra were measured with the following parameters: excitation laser power from 22 $\mu$W to 1560 $\mu$W, acquisition time for each spectrum 3 s, 10 acquisitions per spectrum, with a 100$\times$ objective (NA=0.85, spot size 1~$\mu$m). The maps were collected with the following parameters: excitation laser power 120 $\mu$W, acquisition time for each spectrum 3 s, 4 acquisitions per spectrum, with a 100$\times$ objective (NA=0.85, spot size 1~$\mu$m). A Newport power meter, model 843-R, with a low-power calibrated photodiode sensor was used to measure the incident Raman laser power for different nominal percentages. Atomic force microscopy (AFM) was performed for surface analysis utilizing an Anasys Instruments AFM + microscope system operating in tapping mode. The Gwyddion software package was used to analyze the AFM images.\cite{Nevcas12gwyddion}

\subsection{Computational Details} All calculations were performed with the CP2K\textsuperscript{\cite{2014cp2k,Kuhne20}} program at the DFT level of theory using the Perdew-Burke-Ernzerhof (PBE) exchange and correlation functional.\textsuperscript{\cite{PBE1996}} Second-generation dispersion corrections (D2)\textsuperscript{\cite{Grimme06}} were used to take into account a proper description of van der Waals interactions. Goedecker-Teter-Hutter (GTH) pseudopotentials,\textsuperscript{\cite{GTH1996}} together with double-zeta optimised basis sets (DZVP), were used\cite{vandeV2007}. The Gaussian-and-Plane-Waves (GPW) method as implemented in CP2K\textsuperscript{\cite{2005quickstep}} was used; the energy cutoff for the auxiliary plane-wave basis was set to 340~Ry. The wavefunction convergence criterion was set to $10^{-6}$ Hartree. Geometry optimization of the systems was performed by using the Broyden-Fletcher-Goldfarb-Shanno (BFGS) algorithm by setting a root mean square (RMS) value of $10^{-4}$ Hartree/Bohr for the force and $10^{-4}$ Bohr for the geometry as convergence criteria. The systems were treated as periodic. Functionalized graphene was modeled starting from the hexagonal graphene supercell (a = 1.98 nm) consisting of 154 carbon atoms. After structural minimization, the functionalized system was subject to free molecular dynamics (MD). MD simulations were performed in the NVT ensemble (T~=~300~K) employing a canonical-sampling-through-velocity-escalating (CSVR) thermostat\textsuperscript{\cite{Bussi07}} with a time constant of 500~fs. A single integration time step of 0.4 fs was used. MD simulations were carried out for 38~ps. Vibrational spectra were obtained by calculating the Fourier transform of the atoms velocity auto-correlation function (VACF) taken from the last 36 ps of the simulated trajectory of the system.\textsuperscript{\cite{Thomas13}} Normal modes and vibrational analysis were performed for graphene and functionalized graphene minimized structures using the VIBRATIONAL-ANALYSIS module of CP2K.

%
%
%
%


\phantomsection
\section*{Acknowledgments} 

\addcontentsline{toc}{section}{Acknowledgments} 

The authors thank Fabio Beltram for his continuous support. The authors thank Dr. C. Coletti, from Istituto Italiano di Tecnologia, for the access to the Raman system. This research was partially founded by EU-H2020 FETPROACT LESGO (Agreement No. 952068), and by the Italian Ministry of University and Research (project MONSTRE-2D PRIN2017 KFMJ8E).

\section*{Supporting Information}
Supporting Information is available online or from the authors.


\phantomsection
\bibliographystyle{unsrt}
\bibliography{bib.bib}


\end{document}